\documentclass[prl,reprint,amsfonts,amsmath,amssymb,showpacs,twocolumn]{revtex4}

\usepackage[dvipdfmx]{graphicx}
\usepackage{dcolumn}
\usepackage{bm}
\usepackage{amsthm}
\usepackage{mathrsfs}
\usepackage{comment}
\usepackage{color}

\begin{document}
\title{$\bm{Z}_2$-topology in nonsymmorphic crystalline insulators:
M\"obius twist in surface states}
\author{Ken Shiozaki$^1$}
\author{Masatoshi Sato$^2$}
\author{Kiyonori Gomi$^3$}
\affiliation{
$^1$Department of Physics, Kyoto University,
Kyoto 606-8502, Japan\\
$^2$Department of Applied Physics, Nagoya University,
Nagoya 464-8603, Japan \\ 
$^3$Department of Mathematical Sciences, Shinshu University, 
Nagano, 390-8621, Japan \\ 
}

\date{\today}

\begin{abstract}
It has been known that an anti-unitary symmetry such as time-reversal or
charge conjugation is needed to realize $\bm{Z}_2$ topological phases
in non-interacting systems. 
Topological insulators and superconducting nanowires are
representative examples of such $\bm{Z}_2$ topological matters.
Here we report the first-known $\bm{Z}_2$ topological phase protected
 by only unitary symmetries.    
We show that the presence of a nonsymmorphic space group
 symmetry opens a possibility to realize  $\bm{Z}_2$ topological phases
without assuming any anti-unitary symmetry.
The $\bm{Z}_2$ topological phases are constructed in various dimensions,
 which are closely related to each other by Hamiltonian mapping.
In two and three dimensions, the $\bm{Z}_2$ phases have a surface
consistent with the nonsymmorphic space group symmetry, and thus they
 support topological gapless surface states. 
Remarkably, the surface states have a unique energy dispersion with
 the M\"obius twist, which identifies the $\bm{Z}_2$ phases
 experimentally.  
We also provide the relevant structure in the $K$-theory.  

\end{abstract}
\pacs{}
\maketitle
{\it Introduction.}---
Symmetry is a key for recent developments on
topological phases. 
For instance, time-reversal symmetry and its resultant Kramers
degeneracy are essential
for the stability of quantum spin Hall states \cite{Kane2005,
Bernevig2006} and
three-dimensional (3D) topological insulators \cite{Moore2007, Fu2007a, Roy2009}. 
Also, the particle-hole symmetry (or charge conjugation symmetry) in
superconductors
makes it possible to realize topological superconductors
\cite{Volovik2003, Read2000,Kitaev2001,Sato2003, Fu2008, Qi2009, Roy2008,
Sato2009b, Sato2009c, Sau2010, Sato2009, Sato2010, Fu2010a} which support
exotic Majorana fermions on their
boundary. 
Based on these symmetries,  
many candidate systems for topological insulators and superconductors have
been proposed theoretically and examined experimentally
\cite{Schnyder2008, Hasan2010,Qi2011,TSN2012, Alicea2012, Ando2013}. 

In addition to the general symmetries of time-reversal and
charge-conjugation, materials have their own symmetry specific to the
structures.
In particular, crystals are invariant under space group symmetry, like
inversion, reflection, discrete rotation and so on. 
Such crystalline symmetries also provide a new class of topological
phases, which are dubbed topological crystalline insulators
\cite{Fu2010, Hsieh2012} and
topological crystalline superconductors \cite{MSM2012, Teo2013, Ueno2013, Zhang2013}.
Surface states protected by crystalline symmetry have been confirmed
experimentally\cite{Tanaka2012a,Dziawa2012,Xu2012}. 
Furthermore, a systematic classification of such
topological phases and topological defects has been done theoretically\cite{Chiu2013,Morimoto2013,Shiozaki2014}. 

In the study of topological crystalline insulators and superconductors,  
much attention has been paid for those protected by point group
symmetries\cite{Slager2013, Benalcazar2013,Alexandradinata2014}. 
However, point groups are not only allowed crystalline symmetries.
Space groups contain a transformation which
is not a simple point group operation but 
a combination of a point group operation and a 
nonprimitive lattice transformation.
This class of transformations is called nonsymmorphic.
In spite that many crystals have such nonsymmorphic symmetries, only a
few has been known for their influence on topological phases
\cite{Liu2013a, Parameswaran2013}.  

In this paper, we show that the presence of nonsymmorphic space group
 symmetries provides unique $\bm{Z}_2$ topological phases.
Being different from other known $\bm{Z}_2$ phases, the new $\bm{Z}_2$
 phases need no anti-unitary
 symmetry like time-reversal or charge-conjugation.
We present the $\bm{Z}_2$ topological phases in various dimensions,
 which are closely related to each other.
In two and three dimensions, the $\bm{Z}_2$ phases may have a surface
consistent with the nonsymmorphic space group symmetry, and thus they
 support topological gapless surface states.
Unlike helical surface Dirac modes in other $\bm{Z}_2$ phase, the
 surface states have a unique energy dispersion with M\"obius twist,
 which provides a distinct experimental signal for these phases.  
The $\bm{Z}_2$ topological stability of the surface states and a
 relevant strucuture in the $K$-theory are also discussed.



{\it Nonsymmorphic chiral symmetry in 1D}---
As the simplest example, we first consider a 1D system.
In one-dimension, no nonsymmorphic operation is 
consistent with the existence of a boundary, and thus no boundary zero
energy state 
is topologically protected by this symmetry. 
Nevertheless, we can show that an interesting non-trivial $\bm{Z}_2$ bulk
topological structure appears by a nonsymmorphic unitary symmetry. 
The 1D system is also useful to construct $\bm{Z}_2$ nontrivial topological phases in higher dimensions, which have
gapless boundary states protected by nonsymmorphic symmetries.

\begin{figure}
\includegraphics[width=8.5cm]{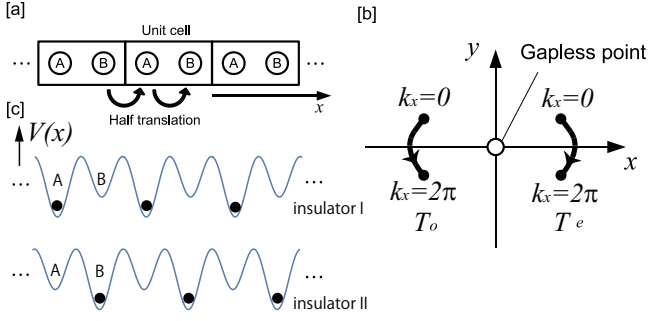} 
\caption{(a) Two inequivalent sites A and B in the unit cell. 
(b)Topologically different trajectories $(x(k_x), y(k_x))$. (c)
 Insulating state I (top) and II (bottom).}
\label{fig:1}
\end{figure}

The symmetry we consider is a nonsymmorphic version of the chiral symmetry:
In stead of the ordinary chiral symmetry,  
\begin{eqnarray}
\{\Gamma, H_{\rm 1D}(k_x)\}=0, 
\quad \Gamma^2=1, 
\end{eqnarray}
where $\Gamma$ is given by a $k_x$-independent unitary matrix, 
we consider a $k_x$-dependent chiral symmetry with
\begin{eqnarray}
\{\Gamma_{\rm 1D}(k_x), H_{\rm 1D}(k_x)\}=0, 
\quad \Gamma_{\rm 1D}^2(k_x)=e^{-ik_x}.
\label{eq:1DNS}
\end{eqnarray}
By imposing $2\pi$-periodicity in $k_x$ on $\Gamma(k_x)$,
the simplest $\Gamma_{\mathrm{1D}}(k_x)$ is
\begin{eqnarray}
\Gamma_{\rm 1D}(k_x)= \begin{pmatrix}
0 & e^{-i k_x} \\
1 & 0
\end{pmatrix},
\end{eqnarray}
where $\Gamma_{\mathrm{1D}}(k_x)$ acts on two inequivalent sites A and B
in the unit cell. 
As illustrated in Fig.\ref{fig:1}(a),
$\Gamma_{\mathrm{1D}}(k_x)$ exchanges these two sites, followed by a {\it half
translation} in the lattice space.

The Hamiltonian with the nonsymmorphic chiral symmetry has a generic form 
\begin{eqnarray}
H_{\rm 1D}(k_x)
=\begin{pmatrix}
x(k_x) & -iy(k_x)e^{-ik_x/2}	\\	 
iy(k_x)e^{ik_x/2} & -x(k_x)
\end{pmatrix}, 
\label{eq:NSchiralHamiltonian}
\end{eqnarray}
with real functions $x(k_x)$ and $y(k_x)$.
The $2\pi$-periodicity of the Hamiltonian, $H_{\mathrm{1D}}(k_x+2\pi)=H_{\mathrm{1D}}(k_x)$, implies 
\begin{eqnarray}
x(k_x+2\pi)=x(k_x), \quad y(k_x+2\pi)=-y(k_x). 
\label{eq:constraint}
\end{eqnarray}
Because the eigenvalues of the Hamiltonian are 
$
E(k_x)=\pm\sqrt{[x(k_x)]^2+[y(k_x)]^2}, 
$
the system is gapped at $E=0$ unless the vector $(x(k_x), y(k_x))$ passes
through the origin $(0,0)$ at some $k_x$.

Now we will show that the Hamiltonian (\ref{eq:NSchiralHamiltonian}) has
two distinct topological phases:
As we show in Fig.\ref{fig:1}(b), 
the Hamiltonian defines a trajectory of $(x(k_x), y(k_x))$
in the $xy$-plane, when $k_x$ changes
from $0$ to $2\pi$.  
From the constraint of Eq.(\ref{eq:constraint}), 
the trajectory forms an open arc, not a closed circle, and  
the end point $(x(2\pi), y(2\pi))$ must be the mirror image
of the start point $(x(0), y(0))$ with respect to the $x$-axis.
The open trajectory passes the $x$-axis odd
number of times. 
More precisely, we have two different ways to across the $x$-axis;
if the trajectory pass the positive
$x$-axis odd (even) number of times, then it must pass the negative
$x$-axis even (odd) number of times. 
See trajectories ${\rm T}_{\rm o}$ and ${\rm T}_{\rm e}$ in
Fig.\ref{fig:1}(c). 
These two different trajectories cannot be continuously
deformed into each other without gap-closing, since they
cannot across the origin without gap-closing, as mentioned in the above. 
Therefore, by counting the parity of times the trajectory passes the
positive $x$-axis, we can identify the two distinct phases of the Hamiltonian
(\ref{eq:NSchiralHamiltonian}).
The $\bm{Z}_2$ nature of the topological phase is discussed in details
in Ref.\cite{SM}.

If the parity is odd (even), then the Hamiltonian is adiabatically
deformed into the $k_x$-independent Hamiltonian $H_{\rm o}$
($H_{\rm e}$) in the below, without gap-closing,
\begin{eqnarray}
H_{\rm o}
=-\sigma_z,
\quad
H_{\rm e}=
\sigma_z,
\end{eqnarray}
with the Pauli matrix $\sigma_i$ $(i=0,x,y,z)$.
These Hamiltonians suggest a simple physical realization of 
the nonsymmorphic chiral symmetry.  
Consider a periodic potential with two different local minima A and
B in the unit cell. See Fig.\ref{fig:1}(c). If the energy of the local
minimum A
(B) is
much higher than B's (A's) and tunnelings between local minima are
neglected, we have an insulating phase I (II) in the half filling, which
effective Hamiltonian is given by $H_{e}$ ($H_o$).
Our argument above implies that these insulating phases are topologically
distinct and they are separated by a topological quantum phase
transition as long as one keeps the symmetry (\ref{eq:1DNS}).
Such a periodic system could be artificially created by cold atoms. 

{\it Nonsymmorphic $Z_2$ symmetry in 2D}---
Much more interesting $\bm{Z}_2$ topological phases protected by
nonsymmorphic symmetries
appear in two and three dimensions. 
In these dimensions, a class of nonsymmorphic symmetries are consistent
with the presence of a surface, 
and thus the symmetry protected gapless edge
states may appear.
Here we present a 2D $\bm{Z}_2$-topological nonsymmorphic insulator,
which supports a unique edge state.


To obtain the $\bm{Z}_2$ phase, we use a Hamiltonian map that increases
the dimension of the system.
This map keeps the topological
structure by shifting symmetries, 
and is known to be useful to classify
the topological (or topological crystalline) insulators/superconductors\cite{Teo2010,Shiozaki2014}. 
In particular, the periodic structure of the topological table is
explained by this map. The details of the map in the present case and
the relevant structure in the K-theory are
given in Ref.\cite{SM}.

From the Hamiltonian mapping, we obtain a representative Hamiltonian of
a 2D $\bm{Z}_2$ topological nonsymmorphic insulator, 
\begin{eqnarray}
H_{\rm 2D}(k_x, k_y)=(m+\cos k_y)\tau_y\otimes H_{\rm 1D}(k_x)+\sin k_y
 \tau_x\otimes\sigma_0, 
\label{eq:2DNSHamiltonian}
\end{eqnarray}
which has a $k_x$-dependent nonsymmorphic symmetry
\begin{eqnarray}
\left[
U(k_x), H_{\rm 2D}(k_x, k_y) \right]=0, 
\quad
U(k_x)=\tau_x\otimes \Gamma_{\rm 1D}(k_x),
\label{eq:NS2D}
\end{eqnarray}
and the additional chiral symmetry, 
\begin{eqnarray}
\{\Gamma, H_{\rm 2D}(k_x, k_y)\}=0, \quad
\Gamma=\tau_z\otimes \sigma_0  
\label{eq:2Dchiral}
\end{eqnarray}
where $\tau_i$ ($i=0,1,2,3$) is the Pauli matrix for the degrees of
freedom on which $\Gamma$ acts. 
These two symmetry operators anticommute
\begin{eqnarray}
\{\Gamma, U_{\rm 2D}(k_x)\}=0. 
\end{eqnarray}
Here note that the nonsymmorphic symmetry $U_{\mathrm{2D}}(k_x)$ commutes
with $H_{\mathrm{2D}}(k_x, k_y)$, although it is constructed from 
$\Gamma_{\mathrm{1D}}(k_x)$ anticommuting with $H_{\mathrm{1D}}(k_x)$.
Whereas any terms consistent with the symmetries (\ref{eq:NS2D})
and (\ref{eq:2Dchiral}) can be added to the Hamiltonian, the basic
topological
properties can be captured by Eq.(\ref{eq:2DNSHamiltonian}).   
For a gapped $H_{\mathrm{1D}}(k_x)$, the system has a gap unless $m=\pm 1$.
Using the symmetries (\ref{eq:NS2D}) and (\ref{eq:2Dchiral}), we can
define a $\bm{Z}_2$ invariant, which is nontrivial (trivial) if
$-1<m<1$ ($m>1$ or $m<-1$)\cite{SM}. 
Without loss of generality, we assume in the following that the parity of
$H_{\mathrm{1D}}(k_x)$ is even, so it is topologically equivalent to $H_{\mathrm{e}}=\sigma_z$.

If we consider a boundary parallel to the $x$-axis, we can keep the
symmetries (\ref{eq:NS2D}) and (\ref{eq:2Dchiral}). 
This boundary supports gapless edge states when
the system is topological ($-1<m<1$):
To demonstrate this, consider a semi-infinite system $(y>0)$ with the
edge at $y=0$.
Since $H_{\mathrm{1D}}(k_x)$ is topologically equivalent to $\sigma_z$, 
we first consider the spacial case of the Hamiltonian
(\ref{eq:2DNSHamiltonian}) with $H_{\mathrm{1D}}(k_x)=\sigma_z$. 
In this particular case, the Hamiltonian $H_{\mathrm{2D}}(k_x, k_y)$ does not
depend on $k_x$, and
thus the topological edge state should be a $k_x$-independent
zero energy state.  
The edge state can be obtained analytically when the system is close to
the topological phase transition at $m=\pm 1$.
Near the topological phase transition, say at $m=-1$, the bulk
gap is nearly closed at $k_y=0$, so the low energy physics is
well-described by the effective Hamiltonian obtained by the expansion of 
Eq.(\ref{eq:2DNSHamiltonian}) around $k_y=0$. 
Then, replacing $k_y$ with
$-i\partial_y$, we have the equation for the edge state
\begin{eqnarray}
\left[
(m+1+\partial_y^2/2)\tau_y\otimes \sigma_z-i\partial_y\tau_x\otimes \sigma_0 
\right]\psi(y)=0,
\end{eqnarray}
with the boundary condition $\psi(0)=0$ and $\psi(\infty)=0$. 
If the system is in the topological side near the transition
i.e. $\delta m\equiv m+1>0$,
the equation has two independent solutions localized at $y=0$
\begin{eqnarray}
&&|\psi_1\rangle=
\begin{pmatrix}
1\\
0 
\end{pmatrix}_{\tau}
\otimes
\begin{pmatrix}
0\\
1 
\end{pmatrix}_{\sigma}
e^{-y}
\sinh\left(\sqrt{-2\delta m+1} y\right)
\nonumber\\
&&|\psi_2\rangle=
\begin{pmatrix}
0\\
1 
\end{pmatrix}_{\tau}
\otimes
\begin{pmatrix}
1\\
0 
\end{pmatrix}_{\sigma}
e^{-y}
\sinh\left(\sqrt{-2\delta m+1} y\right).
\label{eq:edgestate}
\end{eqnarray}
On the other hand, in the  non-topological side ($\delta m<0$), the
solutions are diverge, and the edge states disappear. 
A similar result is found near another transition point at $m=1$. 
%
We have also confirmed numerically the existence of the zero energy edge
mode for the whole region of $-1<m<1$. 

\begin{figure}
\includegraphics[width=9cm]{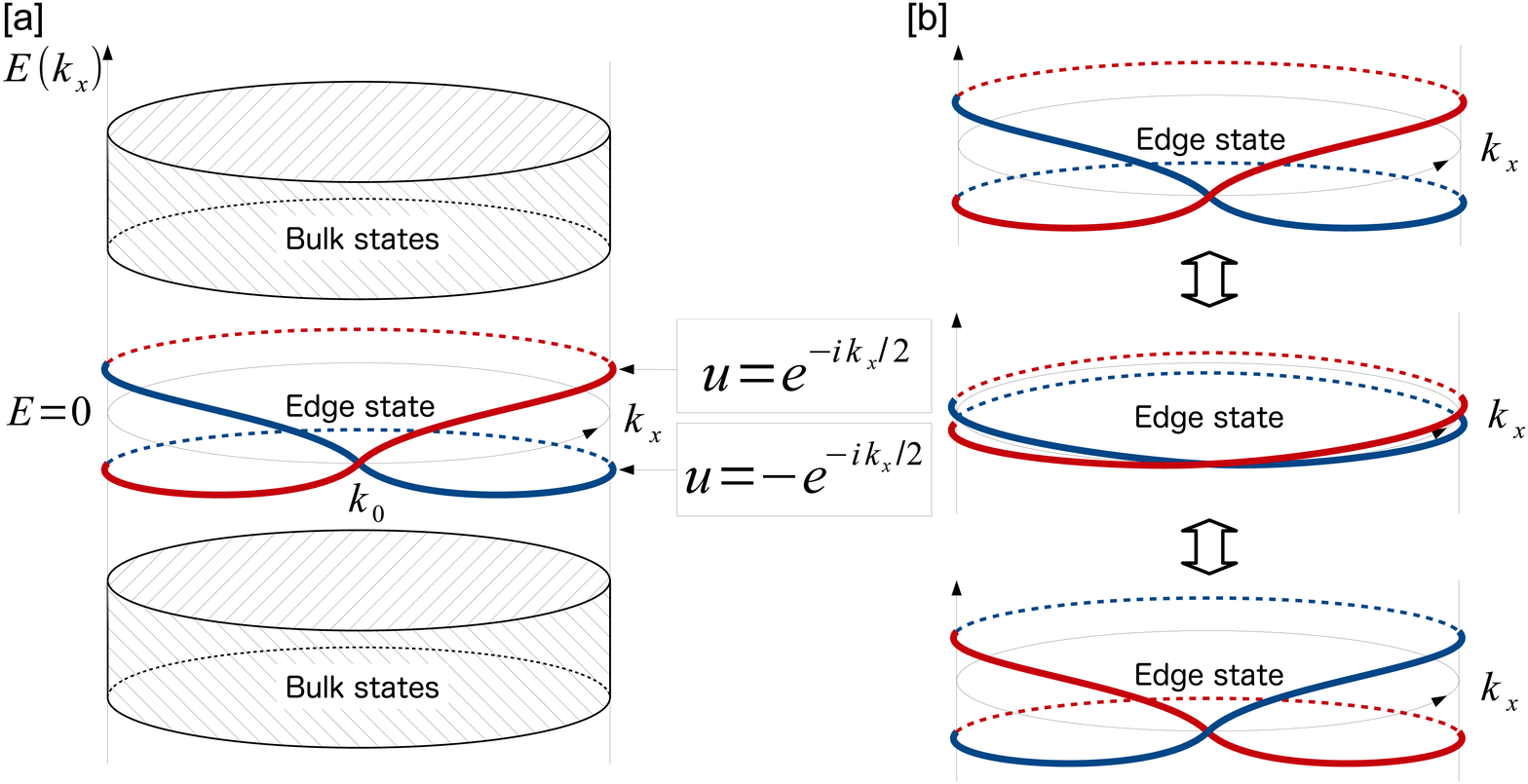} 
\caption{(color online).
Schematic illustration of edge states with M\"obius twist.
(a) The red (blue) line is an edge state in the eigensector of $U(k_x)$ with
 the eigenvalue $u=e^{-ik_x/2}$ ($u=-e^{-ik_x/2}$). (b) An exchange
 process of the eigensectors. }
\label{fig:2DZ2Nonsymmorphic}
\end{figure}

For a general $k_x$-dependent $H_{\mathrm{1D}}(k_x)$, 
the zero energy edge states have a $k_x$-dependent energy dispersion.
By diagonalizing the mixing matrix
$
\langle\psi_i|(\delta m+\partial_y^2/2)\tau_y\otimes (H_{\mathrm{1D}}(k_x)-\sigma_z)|\psi_j\rangle$, 
the energy is evaluated as $E(k_y)\propto\pm y(k_x)$.
Then, from the constraint (\ref{eq:constraint}), there must be an
odd number of zeros for $y(k_x)$ in $k_x\in [0,2\pi]$, and thus
the energy dispersion becomes helical
$E(k_y)\sim \pm c(k_x-k_0)$ around each zero $k_0$,
as illustrated in Fig.\ref{fig:2DZ2Nonsymmorphic} (a).   

Since the Hamiltonian $H_{\mathrm{2D}}(k_x, k_y)$ commutes with $U(k_x)$, 
the helical dispersion is decomposed into chiral and anti-chiral ones, each of
which is an eigenstate of $U(k_x)$.
These two chiral dispersions are mapped to each other 
by the chiral symmetry $\Gamma$, because $\Gamma$ maps a gapless state
to another one, reversing the slope of
the dispersion.
Furthermore, they belong to different eigensectors of
$U(k_x)$, because $\Gamma$ exchanges the eigenvalues of $U(k_x)$ due to
$\{\Gamma, U(k_x)\}=0$.
Therefore, these two chiral dispersions stay gapless without mixing, as
far as the symmetries 
(\ref{eq:NS2D}) and (\ref{eq:2Dchiral}) are retained.


Whereas the above edge state has a similarity to helical edge modes in
quantum spin Hall states, their overall structure in the momentum space
is completely different:
As is seen in Fig.\ref{fig:2DZ2Nonsymmorphic} (a), the present edge
state has a unique energy dispersion with the M\"obius twist, 
which is never seen in other ${\bm Z}_2$ phases.
This twist occurs due to the multivalueness of the eigenvalues $u=\pm
e^{-ik_x/2}$ of $U(k_x)$: 
When one goes round in the $k_x$-direction as  $k_x\rightarrow
k_x+2\pi$, 
$u$ changes the sign, so a chiral dispersion in an eigensector of
$U(k_x)$  turns smoothly into to an anti-chiral one in another
eigensector. 



Another remarkable feature of our edge state 
is that the constituent chiral dispersions can
exchange their eigensectors of $U(k_x)$,  
as illustrated in Fig.\ref{fig:2DZ2Nonsymmorphic} (b).  
This means that any pair of helical dispersions is
topologically unstable:
When a pair of helical dispersions exits,  
we can always realize the situation where a chiral dispersion coexists
with an anti-chiral one in the same eigensector of $U(k_x)$, by
exchanging the eigensectors properly.  
Thus, we can open a gap of helical dispersion by mixing between the chiral and
anti-chiral ones.


The arguments above clearly indicate that helical edge states in this system 
has a $\bm{Z}_2$ stability like helical edge states in quantum spin Hall
systems, although no time-reversal symmetry is required
and the mechanism of the stability is completely different.





{\it Glide reflection symmetry in 3D}---
Finally, we consider the system with glide reflection symmetry,
\begin{eqnarray}
&&G(k_x)H_{\rm 3D}(k_x,k_y,k_z)G^{-1}(k_x)=H_{\rm 3D}(k_x, k_y, -k_z), 
\nonumber\\
&&G^2(k_x)=e^{-ik_x},
\end{eqnarray}
The glide reflection $G(k_x)$ is the combination of reflection with
respect to the $xy$-plane and translation along the $x$-axis by a half
of the lattice spacing.
Since $G^2(k_x)$ results in a translation by a unit
lattice spacing in the $x$-direction, it provides the non-trivial
$e^{-ik_x}$ factor.
The $\bm{Z}_2$ invariant defined by the glide reflection symmetry is
given in Ref.\cite{SM}. 


A representative Hamiltonian with glide reflection symmetry is given by
\cite{SM} 
\begin{eqnarray}
H_{\rm 3D}({\bm k})&=&(m+\cos k_z+\cos k_y)\tau_y\otimes H_{\rm 1D}(k_x) 
\nonumber\\
&+&\sin k_y \tau_x\otimes \sigma_0+\sin k_z \tau_z\otimes \sigma_0,
\nonumber\\
G(k_x)&=&\tau_x\otimes \Gamma_{\rm 1D}(k_x).
\label{eq:3DNSHamiltonian}
\end{eqnarray}
The 3D system is gapped unless $m=\pm 2,0$.
The $\bm{Z}_2$ invariant is non-trivial (trivial) when $-2<m<0$ or $0<m<2$
($m<-2$ or $m>2$) \cite{SM}.

A surface perpendicular to the $y$-axis retains the glide reflection
symmetry, so it may support a gapless surface state protected by this
symmetry. 
For instance, consider a semi-infinite 3D system $(y>0)$ with a
surface at $y=0$, which preserves
the glide reflection symmetry.
In a manner similar to the 2D system, 
for the special but topologically equivalent case with 
$H_{\mathrm{1D}}(k_x)=\sigma_z$, 
we can obtain the surface state
analytically near the topological phase transition at $m=\pm 2$:
For $m\sim -2$, $H_{\mathrm{3D}}(\bm{k})$ is
well approximated by
\begin{eqnarray}
\hat{H}_{\mathrm{3D}}&=&(m+\cos k_z+1-\partial_y^2/2)\tau_y\otimes \sigma_z
\nonumber\\ 
&-&i\partial_y\tau_x\otimes \sigma_0+\sin k_z \tau_z\otimes \sigma_0.
\end{eqnarray}
We find that $|\psi_1\rangle$ and $|\psi_2\rangle$ in
Eq.(\ref{eq:edgestate}) with $\delta m=m+\cos k_z+1$ satisfy the
Schr\"odinger equation,
\begin{eqnarray}
\hat{H}_{\rm 3D}\psi_i(y)=E_i(k_z) \psi_i(y),
\end{eqnarray}
with $E_1(k_z)=\sin k_z$ and $E_2(k_z)=-\sin k_z$, respectively. 
When the system is in the topological side near the transition,
i.e. $-2<m<0$, $\delta m$ is positive (negative) at $k_z=0$ ($k_z=\pi$). 
Thus, they meet the boundary condition $\psi_{i}(0)=0$ and
$\psi_{i}(\infty)=0$ near $k_z=0$, while they diverge near $k_z=\pi$.
This means that they form surface states with the linear dispersion
$E(k_z)=\pm k_z$ near $k_z=0$, which merge into
bulk states near $k_z=\pi$.
On the other hand, in the topologically trivial side, i.e. $m<-2$,
$\delta m$ is always negative, so $|\psi_1\rangle$ and $|\psi_2\rangle$
are no longer physical states anymore. 
A similar analysis works for $0<m<2$, although the surface states appear
near $k_z=\pi$ in this case.  

\begin{figure}
\includegraphics[width=8.5cm]{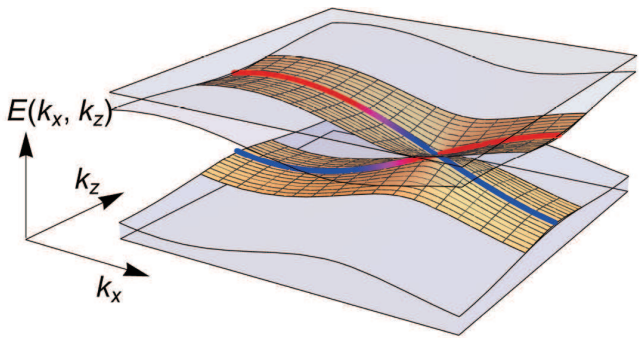} 
\caption{(color online). A surface state protected by glide reflection
 symmetry. The spectrum at $k_z=\Lambda$ has a M\"obius twist in the
 $k_x$-direction: Along the $k_x$-direction, the red branch with the eigenvalue $e^{ik_x/2}$ of
 $G(k_x)$ turns into the blue one with the eigenvalue $-e^{ik_x/2}$.}
\label{fig:3Dglide} 
\end{figure}

For a general $H_{\mathrm{1D}}(k_x)$, the surface states have a dispersion in
the $k_x$-direction, as well as in the $k_z$-direction:
Like the 2D case, 
the two surface modes, $|\psi_1\rangle$ and $|\psi_2\rangle$, are mixed.
The spectrum of the surface states becomes
$E(k_x,k_z)=\pm \sqrt{cy^2(k_x)+\sin^2 k_z}$.  
($c$ is a constant.)
From the constraint (\ref{eq:constraint}), $y(k_x)$ has an odd
number of zeros, and thus 
the surface states have the corresponding odd
number of Dirac cones in the spectrum,
as illustrated in Fig.\ref{fig:3Dglide}.

In the glide invariant plane at $k_z=\Lambda$ ($\Lambda=0, \pi$) in the
Brillouin zone, the Dirac cone
has helical dispersions $E\sim \pm c(k_x-k_0)$ in the $k_x$-direction.
Since $H_{\rm 3D}({\bm k})$ commutes with $G(k_x)$ at $k_z=\Lambda$,  
the helical dispersion can be divided into
two eigensectors of $G(k_x)$, which have chiral dispersion and
anti-chiral dispersion, respectively.
These two chiral dispersions cannot mix, so a single
Dirac cone is topologically stable.
On the other hand, a pair of Dirac cone is topologically unstable: 
From a process similar to Fig.\ref{fig:2DZ2Nonsymmorphic} (b), 
the eigensectors can exchange without gap-closing.
Therefore, from a similar argument in the 2D case, helical
dispersions for a pair of Dirac cones can be gapped. 

As in the 2D case, the obtained surface state has a very unique
feature: In the $k_x$-direction, which is the direction of the translation
for the glide, the surface state has an energy dispersion with the
M\"obius twist. 
Furthermore, along the same direction, 
the surface state is detached from the bulk
spectrum. 
Indeed, by adiabatically changing $H_{\mathrm{1D}}(k_x)$ as 
$H_{\mathrm{1D}}=\sigma_z$, the surface state
becomes completely flat at $E=0$ in the $k_x$-direction. 
This feature is never seen in surface Dirac modes in other ${\bm Z}_2$
phases. 
Any stable Dirac mode in
other ${\bm Z}_2$ phases
bridge the bulk
conduction and valence bands in any direction in the surface Brillouin
zone.  
This remarkable feature in the spectrum can be detected by angle-resolved
photoemission, which provides a distinct evidence of this novel
$\bm{Z}_2$ phase.

{\it Summary}---
We have revealed that nonsymmorphic crystalline symmetries such as
glide reflection symmetry
provide a class of novel ${\bm Z}_2$ phases. 
They are related to each other by Hamiltonian mapping, which is
justified by the K-theory \cite{SM}. 
These ${\bm Z}_2$ phases predict remarkable surface states that
have the M\"obius twist in the spectrum, which can be detectable
experimentally.

{\it Note added}---
After this work was finalized, there appeared a complementary and
independent work  \cite{Fang2015} which has some overlap with our results.

M.S is supported by the JSPS
(No.25287085) and KAKENHI
Grants-in-Aid (No.22103005) from MEXT.
K.S. is supported by a JSPS Fellowship for Young Scientists, and 
K.G. is supported by the Grant-in-Aid for Young Scientists (B 23740051),
JSPS.

\bibliography{Z2NonsymmorphicTCI}

\appendix


\setcounter{equation}{0}
\renewcommand{\theequation}{S\arabic{equation}}

\vspace*{1.5ex}
\centerline{\large \bf Supplementary Material}

\section{Hamiltonian mapping}

Here we introduce a Hamiltonian mapping which relates topological phase in
different dimensions.
A similar map has been used in the classification of topological
insulators and superconductors defined on a sphere ${\bm k}\in S^d$ in
the momentum space\cite{Teo2010,Shiozaki2014}. 
We generalize the idea to insulators with nonsymmorphic symmetries.

First we review the Hamiltonian mapping used in topological insulators
and superconductors.
The map is given as follows:
If a Hamiltonian $H({\bm k})$ on a $d$-dimensional sphere
${\bm k}\in S^{d}$ has chiral symmetry,
$
\{
\Gamma, H({\bm k}) 
\}=0, 
$
with the chiral operator  $\Gamma$,
then the map is
\begin{eqnarray}
H({\bm k},\theta)=\sin \theta H({\bm k})+\cos \theta \Gamma,  
\quad \theta\in[0,\pi],
\label{seq:raising1}
\end{eqnarray}
and if not, it is 
\begin{eqnarray}
H({\bm k}, \theta)=\sin \theta \tau_y \otimes H({\bm k})+ \cos \theta
 \tau_x\otimes{\bm 1},
\,\, \theta\in[0,\pi],
\label{seq:raising2}
\end{eqnarray}
where ${\bm 1}$ is the unit matrix with the same dimension as $H({\bm k})$.
Since the mapped Hamiltonian $H({\bm k},\theta)$ is independent of ${\bm
k}\in S^d$ at $\theta=0$ and $\pi$, the base space $({\bm k}, \theta)\in
S^{d}\times [0,\pi]$ of $H({\bm k}, \theta)$ can be regarded as a
$(d+1)$-dimensional sphere $S^{d+1}$ by shrinking $S^d$ to a point at
$\theta=0$ and $\pi$, respectively. 
Thus the mapped Hamiltonian $H({\bm k}, \theta)$ is defined on  $S^{d+1}$.
Furthermore, it can be shown that the map is isomorphic and thus the original
$H({\bm k})$ and the mapped $H({\bm k},\theta)$ have the same
topological structures.
This map relates topological insulators in different dimensions, and it
enables us to study their topological phases systematically.

Using the above isomorphic map, we can construct the 2D
insulator
\begin{eqnarray}
H_{\rm 2D}(k_x, k_y)=(m+\cos k_y)\tau_y\otimes H_{\rm 1D}(k_x)
\nonumber\\
+\sin k_y\tau_x\otimes\sigma_0, 
\label{seq:2DNSHamiltonian}
\end{eqnarray}
with symmetries
\begin{eqnarray}
&&\left[
U(k_x), H_{\rm 2D}(k_x, k_y) \right]=0, 
\quad
U(k_x)=\tau_x\otimes \Gamma_{\rm 1D}(k_x),
\nonumber\\
&&\{\Gamma, H_{\rm 2D}(k_x, k_y)\}=0,
\quad
\Gamma=\tau_z\otimes \sigma_0,
\label{seq:NS2D}
\end{eqnarray}
which is topologically nontrivial (trivial) for $-1<m<1$ ($m>1$ or $m<-1$).

\begin{figure}
\includegraphics[width=8cm]{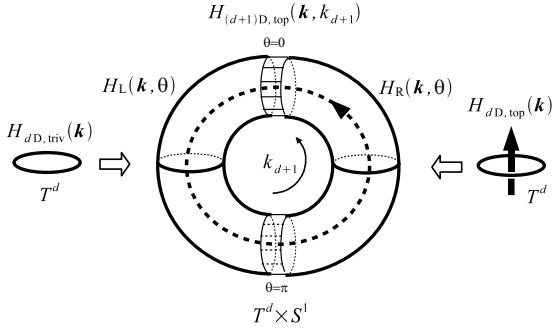} 
\caption{Hamiltonian mapping. Two Hamiltonians $H_{\rm R}({\bm k},
 \theta)$ and $H_{\rm R}({\bm k}, \theta)$ defined on $T^d\times
 [0,\pi]$ are sewn at $\theta=0$ and $\pi$.}
\label{sfig:Hamiltonian_mapping}
\end{figure}

The basic idea is as follows: For $H_{\rm 1D}({\bm k})$ on $k_x\in
S^1$, consider the following two Hamiltonians defined on $(k_x,\theta)\in S^1\times
[0,\pi]$, 
\begin{eqnarray}
&&H_{\rm R}(k_x,\theta)=\sin \theta \tau_y\otimes H_{\rm 1D}(k_x)+\cos\theta
 \tau_x\otimes \sigma_0,
\nonumber\\ 
&&H_{\rm L}(k_x,\theta)=\sin \theta \tau_y\otimes [-H_{\rm 1D}(k_x)]+\cos\theta
 \tau_x\otimes \sigma_0,
\end{eqnarray}
which are obtained by the isomorphic map (\ref{seq:raising2}).
They have the symmetry,
\begin{eqnarray}
[U(k_x), H_{{\rm R},{\rm L}}(k_x, \theta)]=0,
\quad
\{
\Gamma, H_{{\rm R},{\rm L}}(k_x, \theta)
\}
=0,
\label{seq:2DNStheta}   
\end{eqnarray}
with $U(k_x)=\tau_x\otimes\Gamma_{\rm 1D}(k_x)$ and
$\Gamma=\tau_z\otimes\sigma_0$.  
Since $H_{\rm 1D}(k_x)$ and $-H_{\rm 1D}(k_x)$ have different ${\bm
Z}_2$ numbers, either $H_{\rm R}(k_x, \theta)$ or $H_{\rm L}(k_x,
\theta)$, but not both is topologically nontrivial. 
These two Hamiltonians coincide at $\theta=0$ and $\pi$, respectively.
Thus, by sewing these two Hamiltonians at $\theta=0$ and $\pi$, as
illustrated in Fig.\ref{sfig:Hamiltonian_mapping}, we can obtain a system defined on a
two-dimensional torus $T^2$.
The resultant system has a non-trivial $\bm{Z}_2$ number, which is
obtained as the total $\bm{Z}_2$ numbers of $H_{\rm R}(k_x,
\theta)$ and $H_{\rm L}(k_x, \theta)$.  

To obtain an explicit Hamiltonian of the system on $T^2$, we change the
variable $\theta$ as
$\theta=\pi/2-k_y$ in $H_{\rm R}(k_x,\theta)$    
and $\theta=k_y-\pi/2$ in $H_{\rm L}(k_x,\theta)$,
respectively.  
For the new variable, $H_{\rm R}$ and $H_{\rm L}$ have the same form
as $H_{\rm 2D}(k_x, k_y)$
\begin{eqnarray}
H_{\rm 2D}(k_x, k_y)=\cos k_y \tau_y\otimes H_{\rm 1D}(k_x)+\sin k_y
 \tau_x\otimes \tau_0, 
\label{seq:2DHamiltonian}
\end{eqnarray}
where $H_{\rm L}$ and $H_{\rm R}$ are smoothly connected at $k_y=\pi/2$
and $k_y=-\pi/2$, respectively.
Equation (\ref{seq:2DHamiltonian}) is the Hamiltonian of the
sewn system.  
Note that we may adiabatically add a term preserving the symmetries
(\ref{seq:2DNStheta}) to the Hamiltonian 
without changing its topological property 
unless the bulk gap of the system closes.
Thus we can finally modify
(\ref{seq:2DHamiltonian}) in the form of Eq.(\ref{seq:2DNSHamiltonian})
with $-1<m<1$.

In a similar manner, we can obtain a system on $T^2$ with trivial ${\bm
Z}_2$ topology. In this case, we use the same Hamiltonian for $H_{\rm
R}(k_x, \theta)$ and $H_{\rm L}(k_x, \theta)$, 
\begin{eqnarray}
&&H_{\rm R}(k_x, \theta)=H_{\rm L}(k_x,\theta)
\nonumber\\
&&=\sin \theta \tau_y\otimes H_{\rm 1D}(k_x)+\cos\theta
 \tau_x\otimes \sigma_0, 
\label{seq:2DLR2}
\end{eqnarray}
with $\theta\in [0,\pi]$.
Even when $H_{\rm R}$ and $H_{\rm L}$ have non-trivial $\bm{Z}_2$
numbers, they are canceled
by sewing them at $\theta=0$ and $\theta=\pi$.
An explicit form of the sewn Hamiltonian is obtained as follows.
Because $\sin \theta\geq 0$, we can add a positive constant $m$ to $\sin
\theta$ in Eq.(\ref{seq:2DLR2}) without gap-closing,
\begin{eqnarray}
&&H_{\rm R}(k_x, \theta)=H_{\rm L}(k_x,\theta)
\nonumber\\
&&=(m+\sin \theta) \tau_y\otimes H_{\rm 1D}(k_x)+\cos\theta
 \tau_x\otimes \sigma_0, 
\end{eqnarray}
where we gradually increase $m$ as it satisfies $m>1$. 
Then we can adiabatically change the coefficient of $\sin \theta$ in
$H_{\rm L}(k_x,\theta)$ as $\sin \theta \rightarrow -\sin \theta$,
without gap-closing. 
As a result, $H_{\rm R}$ and $H_{\rm L}$ can be 
\begin{eqnarray}
H_{\rm R}(k_x, \theta)
=(m+\sin \theta) \tau_y\otimes H_{\rm 1D}(k_x)+\cos\theta
 \tau_x\otimes \sigma_0, 
\nonumber\\
H_{\rm L}(k_x, \theta)
=(m-\sin \theta) \tau_y\otimes H_{\rm 1D}(k_x)+\cos\theta
 \tau_x\otimes \sigma_0, 
\end{eqnarray}
with $m>1$.
Finally, by changing the
variable $\theta$ as
$\theta=\pi/2-k_y$ in $H_{\rm R}(k_x,\theta)$    
and $\theta=k_y-\pi/2$ in $H_{\rm L}(k_x,\theta)$,
respectively,   
we find that $H_{\rm R}$ and $H_{\rm L}$ have the form of
Eq.(\ref{seq:2DNSHamiltonian}) with $m>1$, where $H_{\rm R}$ and $H_{\rm
L}$ are smoothly sewn up at $k_y=\pm \pi/2$.
We note that if we take the starting Hamiltonians as
\begin{eqnarray}
&&H_{\rm R}(k_x, \theta)=H_{\rm L}(k_x,\theta)
\nonumber\\
&&=\sin \theta \tau_y\otimes[-H_{\rm 1D}(k_x)]+\cos\theta
 \tau_x\otimes \sigma_0, 
\end{eqnarray}
we can obtain Eq.(\ref{seq:2DNSHamiltonian}) with $m<-1$, in a similar manner.

The same idea is available to obtain the 3D insulators
\begin{eqnarray}
H_{\rm 3D}({\bm k})&=&(m+\cos k_z+\cos k_y)\tau_y\otimes H_{\rm 1D}(k_x) 
\nonumber\\
&+&\sin k_y \tau_x\otimes \tau_0+\sin k_z \tau_z\otimes \tau_0,
\label{seq:3DNSHamiltonian}
\end{eqnarray}
with the glide reflection symmetry,
\begin{eqnarray}
&&G(k_x)H_{\rm 3D}(k_x,k_y,k_z)G^{-1}(k_x)=H_{\rm 3D}(k_x, k_y, -k_z), 
\nonumber\\
&&G(k_x)=\tau_x\otimes \Gamma_{\rm 1D}(k_x), 
\end{eqnarray}
which is $\bm{Z}_2$-non-trivial ($\bm{Z}_2$-trivial) for $-2<m<0$ or
$0<m<2$ ($m<-2$ or $m>2$): Since $H_{\rm 2D}$ in
Eq.(\ref{seq:2DNSHamiltonian}) is chiral
symmetric, we use the isomorphic map (\ref{seq:raising1}) to have $H_{\rm
R}(k_x, k_y, \theta)$ and $H_{\rm L}(k_x, k_y, \theta)$, 
\begin{eqnarray}
H_{\rm R}(k_x,k_y, \theta)=\sin \theta H_{\rm 2D}(k_x, k_y)+\cos\theta
 \Gamma, 
\nonumber\\
H_{\rm L}(k_x,k_y, \theta)=\sin \theta H'_{\rm 2D}(k_x, k_y)+\cos\theta
 \Gamma,  
\end{eqnarray}
where we denote $H_{\rm 2D}$ in $H_{\rm L}$ as $H_{\rm 2D}'$ as it can
be different from $H_{\rm 2D}$ in $H_{\rm R}$.
$H_{\rm R}$ and $H_{\rm L}$ have the same $\bm{Z}_2$ topological number
as $H_{\rm 2D}$ and $H_{\rm 2D}'$, respectively.
By jointing $H_{\rm R}$ and $H_{\rm L}$ at $\theta=0$ and
$\pi$, we can have a Hamiltonian $H_{\rm 3D}$ defined on a 3D torus $T^3$.
If either $H_{\rm R}$ or $H_{\rm L}$, but not both is ${\bm
Z}_2$-nontrivial, $H_{\rm 3D}$ is $\bm{Z}_2$-nontrivial. In other
cases, $H_{\rm 3D}$ is $\bm{Z}_2$-trivial.
Then, one can show that with a suitable adiabatic deformation, $H_{\rm
3D}$ takes the form of Eq.(\ref{seq:3DNSHamiltonian}) without
gap-closing.   

\section{$\bm{Z}_2$ invariants for nonsymmorphic systems}

\subsection{1D case}
Here we generalize the $\bm{Z}_2$ invariant defined for the simplest
$2\times 2$ Hamiltonian
(\ref{eq:NSchiralHamiltonian}) in the main text, to that for the general
Hamiltonian.  

The nonsymmorphic chiral symmetry is given by
\begin{eqnarray}
\{\Gamma_{1D}(k_x), H_{\rm 1D}(k_x)\}=0,
\quad \Gamma^2_{\rm 1D}(k_x)=e^{-ik_x}.
\label{seq:1DNSHamiltonian}
\end{eqnarray}
By imposing $2\pi$-periodicity in $k_x$ on $\Gamma_{\rm 1D}(k_x)$, a
general form of $\Gamma_{\rm 1D}(k_x)$ is given by
\begin{eqnarray}
\Gamma_{\rm 1D}(k_x)=
\begin{pmatrix}
0 & e^{-ik_x} \\
1 & 0
\end{pmatrix}
\otimes {\bm 1}_{N\times N}, 
\label{seq:1DNSmatrix}
\end{eqnarray}
with the $N\times N$ unit matrix ${\bm 1}_{N\times N}$.
In this basis, the Hamiltonian $H_{\rm 1D}(k_x)$ with the nonsymmorphic
chiral symmetry takes the form
\begin{eqnarray}
H_{\rm 1D}(k_x)=
\begin{pmatrix}
X(k_x) & -iY(k_x)e^{-ik_x/2} \\
i Y(k_x) e^{ik_x/2} & -X(k_x)
\end{pmatrix}, 
\end{eqnarray}
where $X(k_x)$ and $Y(k_x)$ are $N\times N$ hermitian matrices.
Since $H_{\rm 1D}(k_x)$ is $2\pi$-periodic in $k_x$, 
$X(k_x)$ and $Y(k_x)$ satisfy
\begin{eqnarray}
X(k_x+2\pi)=X(k_x),
\quad
Y(k_x+2\pi)=-Y(k_x).
\end{eqnarray}

Now we introduce 
the following $N\times N$ matrix $Z(k_x)$ 
\begin{eqnarray}
Z(k_x)=X(k_x)+iY(k_x), 
\end{eqnarray}
which has the constraint
\begin{eqnarray}
Z(k_x+2\pi)=Z^{\dagger}(k_x). 
\label{seq:1Dconstraint}
\end{eqnarray}
Because one can prove the relation
\begin{eqnarray}
{\rm det}H_{\rm 1D}(k_x)=|{\rm det}Z(k_x)|^2, 
\label{seq:relation}
\end{eqnarray}
${\rm det}Z(k_x)\neq 0$ when $H_{\rm 1D}(k_x)$ is gapped at $E=0$ 
(namely, when ${\rm det}H_{\rm 1D}(k_x)\neq 0$).

Denoting the real and imaginary parts of ${\rm det}Z(k_x)$ as $x(k_x)$
and $y(k_x)$, respectively, the relation (\ref{seq:relation}) implies 
$x^2(k_x)+y^2(k_x)\neq 0$ for a gapped $H_{\rm 1D}(k_x)$.
Furthermore, from Eq.(\ref{seq:1Dconstraint}), we have
\begin{eqnarray}
x(k_x+2\pi)=x(k_x),
\quad
y(k_x+2\pi)=-y(k_x). 
\end{eqnarray}
Since $x(k_x)$ and $y(k_x)$ defined here have the same property as
those in the main text,
we can define the $\bm{Z}_2$ invariant in the same manner.

As is shown in the main text, the simplest Hamiltonian with
the non-trivial ${\bm Z}_2$ invariant is $H_{\rm o}=-\sigma_z$,
which gives $(x(k_x), y(k_x))=(-1,0)$.
To confirm the ${\bm Z}_2$ nature, consider the direct sum   
$H_{\rm o}\oplus H_{\rm o}$.
In the basis where $\Gamma_{\rm 1D}(k_x)$ takes the form of
Eq.(\ref{seq:1DNSmatrix}), $H_{\rm o}\oplus H_{\rm o}$
gives $X(k_x)=-{\bm 1}_{2\times 2}$ and $Y(k_x)=0$.
Thus, we find $(x(k_x), y(k_x))=(1,0)$ for $H_{\rm o}\oplus H_{\rm o}$,
which implies that $H_{\rm o}\oplus H_{\rm o}$ is ${\bm Z}_2$-trivial.

\subsection{2D case}

In this section, we define the $\bm{Z}_2$ invariant for the 2D Hamiltonian  
which has the nonsymmorphic symmetry
\begin{eqnarray}
\left[
U(k_x), H_{\rm 2D}(k_x, k_y) \right]=0, 
\quad
U(k_x)^2=e^{-ik_x},
\label{seq:NS2D}
\end{eqnarray}
as well as the ordinary chiral symmetry,
\begin{eqnarray}
\{H_{\rm 2D}(k_x, k_y), \Gamma\}=0,
\quad \Gamma^2=1. 
\end{eqnarray}
These symmetries are anticommute,
\begin{eqnarray}
\{\Gamma, U(k_x)\}=0. 
\label{seq:anticommutation}
\end{eqnarray}

Consider the Schr\"odinger equation given by
\begin{eqnarray}
H_{\rm 2D}(k_x, k_y)|u_n(k_x,k_y)\rangle=E_n(k_x,k_y)|u_n(k_x,k_y)\rangle, 
\end{eqnarray}
where $n$ is the band index. 
We assume that the system is gapped at $E=0$, and the Fermi energy is
inside the gap.
It is convenient here to use a positive (negative) $n$ to represent a
positive (negative) energy band. 

Since $H_{\rm 2D}(k_x, k_y)$ commutes with
$U(k_x)$, the solution $|u_n(k_x,k_y)\rangle$ are taken as eigenstates
of $U(k_x)$ 
\begin{eqnarray}
U(k_x)|u_n^{\pm}(k_x,k_y)\rangle=\pm e^{-ik_x/2}|u_n^{\pm}(k_x,k_y)\rangle. 
\end{eqnarray}
The chiral symmetry implies that 
if $|u_n^{\pm}(k_x,k_y)\rangle$ is a positive (negative) energy band,
$\Gamma|u_n^{\pm}(k_x,k_y)\rangle$ is a negative (positive) energy band.
From the anticommutation relation (\ref{seq:anticommutation}), it is
also found that
$\Gamma|u_n^{\pm}(k_x,k_y)\rangle$ is an eigenstate of $U(k_x)$ with
the eigenvalue $\mp e^{ik_x/2}$.
Therefore, we can place the relation
\begin{eqnarray}
|u_n^{\pm}(k_x,k_y)\rangle =\Gamma|u_{-n}^{\mp}(k_x,k_y)\rangle. 
\label{seq:pm}
\end{eqnarray}

A key character of the nonsymmorphic symmetry $U(k_x)$ is
that its eigenvalues $\pm e^{-ik_x/2}$ do not have the same
periodicity as $U(k_x)$ itself: They change their sign when $k_x\rightarrow
k_x+2\pi$. 
As a result, $|u_n^{+}(k_x, k_y)\rangle$ and
$|u_n^{-}(k_x+2\pi, k_y)\rangle$ have the same eigenvalue of $U(k_x)$,
satisfying the same Schr\"odinger equation.
Thus, they are the same state up to a $U(1)$ gauge factor,
\begin{eqnarray}
|u_n^{+}(k_x+2\pi,k_y)\rangle=e^{i\theta_{n}(k_x,k_y)}
|u_n^{-}(k_x, k_y)\rangle.
\label{seq:twist} 
\end{eqnarray}
This relation gives a non-trivial relation in Berry phases:
Introducing the gauge field in the momentum space,
\begin{eqnarray}
A_i^{\pm}(k_x,k_y)
=i \sum_{n<0}\langle u_n^{\pm}(k_x,k_y)
|\partial_{k_i}u_n^{\pm}(k_x,k_y)\rangle,   
\end{eqnarray}
we define the Berry phases $\gamma^{\pm}(k_x)$ as
\begin{eqnarray}
e^{i\gamma^{\pm}(k_x)}
=\exp\left(
i\oint dk_y A_y^{\pm}(k_x, k_y) 
\right).
\end{eqnarray}
Since Eq.(\ref{seq:twist}) implies
\begin{eqnarray}
A_i^{\pm}(k_x+2\pi, k_y)=A_i^{\mp}(k_x,k_y)
-\sum_{n<0}\partial_{k_i}\theta_n(k_x, k_y),
\end{eqnarray} 
the Berry phases satisfy
\begin{eqnarray}
e^{i\gamma^{+}(k_x+2\pi)}=e^{i\gamma^{-}(k_x)}
e^{-i\oint dk_y \sum_{n<0}\partial_{k_y}\theta_n(k_x,k_y)}. 
\end{eqnarray}
From the periodicity in $k_y$, the integral $\oint dk_y
\partial_{k_y}\theta_n$ should be $2\pi N_n$ with a integer $N_n$, and
thus we have
\begin{eqnarray}
e^{i\gamma^{+}(k_x+2\pi)}=e^{i\gamma^{-}(k_x)}.
\label{seq:gammarelation}
\end{eqnarray}

Now we use Eq.(\ref{seq:pm}). This equation implies
\begin{eqnarray}
&&A_i^+(k_x,k_y)+A_i^-(k_x,k_y)
\nonumber\\
&&=i\sum_n\langle u_n^+(k_x, k_y)
|\partial_{k_i} u_n^+(k_x,k_y)\rangle,  
\end{eqnarray}
where the summation in the right hand side is taken for all $n$.
Therefore, from the completeness relation, we find that $A_i^{+}(k_x,
k_y)+A_i^{-}(k_x, k_y)$ is a total derivative of a function, which yields 
\begin{eqnarray}
e^{i[\gamma^{+}(k_x)+i\gamma^{-}(k_x)]}=1. 
\end{eqnarray}
Combining this with Eq.(\ref{seq:gammarelation}), we finally have
\begin{eqnarray}
e^{i\gamma^{+}(k_x+2\pi)}=e^{-i\gamma^{+}(k_x)}. 
\label{seq:2Dconstraint}
\end{eqnarray}

Using this relation, we can define the $\bm{Z}_2$ invariant in the same
manner as the 1D case:
Denoting the real and imaginary parts of
$e^{i\gamma^+(k_x)}$ as $x(k_x)$ and $y(k_x)$, respectively,
we can introduce a nonzero two-dimensional vector $(x(k_x), y(k_x))$.
Then Eq.(\ref{seq:2Dconstraint}) gives the constraint
\begin{eqnarray}
x(k_x+2\pi)=x(k_x),
\quad
y(k_x+2\pi)=-y(k_x),
\end{eqnarray}
which is exactly the same as Eq.(\ref{eq:constraint}). 
Therefore, if the trajectories
$(x(k_x), y(k_x))$ passes the positive $x$-axis odd (even) number of times, 
the system is topologically non-trivial (trivial).

The ${\bm Z}_2$ invariant of the Hamiltonian (\ref{eq:2DNSHamiltonian})
is evaluated as follows. 
It is sufficient to consider the case
with $H_{\rm 1D}(k_x)=\sigma_z$
since $H_{\rm 1D}(k_x)$ can deform into 
$\sigma_z$ without gap-closing. 
$H_{\rm 2D}(k_x, k_y)$ is block
diagonal in the diagonal basis of $U(k_x)$,  
and in the sector with the eigenvalue $u=\pm
e^{-ik_x/2}$ of $U(k_x)$, 
it is given by
\begin{eqnarray}
H_{\rm 2D}^{\pm}=\pm
\begin{pmatrix}
\sin k_y & i(m+\cos k_y) \\
-i(m+\cos k_y) & -\sin k_y 
\end{pmatrix}.
\end{eqnarray}
From this, we obtain
\begin{eqnarray}
\gamma^{\pm}(k_x)=\left\{
\begin{array}{ll}
0, 
& \mbox{for $m<-1$}\\
\pi, 
&\mbox{for $-1<m<1$}\\
0,
& \mbox{for $m>1$}
\end{array}
\right. ,
\end{eqnarray}
which implies the ${\bm Z}_2$ invariant is non-trivial (trivial) if
$-1<m<1$ ($m>1$ or $m<-1$). 

\subsection{3D case}

Finally, we define the $\bm{Z}_2$ topological invariant associated with
glide symmetry
\begin{eqnarray}
&&G(k_x)H_{\rm 3D}(k_x, k_y, k_z)G^{-1}(k_x)=H_{\rm 3D}(k_x,k_y,-k_z ),
\nonumber\\
&&G^2(k_x)=e^{-ik_x}. 
\end{eqnarray}

From solutions of the Schr\"odinger equation 
\begin{eqnarray}
H_{\rm 3D}({\bm k})|u_n({\bm k})\rangle=E_n({\bm k})|u_n({\bm k})\rangle, 
\end{eqnarray}
we introduce the gauge field $A_i({\bm k})$ in the momentum space,
\begin{eqnarray}
A_i({\bm k})=i\sum_{E_n({\bm k})<E_{\rm F}}
\langle u_n({\bm k})|\partial_{k_i}u_n({\bm k})\rangle, 
\end{eqnarray}
where $E_{\rm F}$ is the Fermi energy.
On the glide invariant plane at $k_z=\Lambda$ ($\Lambda=0,\pi$), 
the glide operator $G(k_x)$ commutes with $H_{\rm 3D}$,
\begin{eqnarray}
[G(k_x), H_{\rm 3D}(k_x,k_y, \Lambda)]=0, 
\end{eqnarray}
and thus the solutions of the Schr\"odinger equation are simultaneously eigenstates 
of $G(k_x)$, 
\begin{eqnarray}
G(k_x)|u_n^{\pm}(k_x, k_y, \Lambda)\rangle
=\pm e^{-ik_x/2} |u_n^{\pm}(k_x, k_y, \Lambda)\rangle.
\end{eqnarray}
Correspondingly, we can decompose $A_i({\bm k})$ into two parts,
\begin{eqnarray}
A_i(k_x,k_y,\Gamma)=A_i^{+}(k_x,k_y,\Lambda)+A_i^{-}(k_x,k_y,\Lambda) 
\end{eqnarray}
with
\begin{eqnarray}
&&A_i^{\pm}(k_x, k_y, \Lambda)
\nonumber\\
&&=i\sum_{E_n<E_{\rm F}}
\langle u_n^{\pm}(k_x,k_y, \Lambda)|\partial_{k_i}u_n^{\pm}(k_x, k_y, \Lambda)\rangle. 
\end{eqnarray}
In a manner similar to $U(k_x)$ in 2D, the eigenvalues of $G(k_x)$ do
not have the same periodicity in $k_x$ as $G(k_x)$ itself, and they change
their sign when $k_x\rightarrow k_x+2\pi$. 
As a result, we have a twisted boundary condition,
\begin{eqnarray}
|u_n^{\pm}(k_x+2\pi, k_y, \Lambda)\rangle
=e^{i\theta^{\pm}_n(k_x, k_y, \Lambda)} |u_n^{\pm}(k_x, k_y, \Lambda)\rangle,
\end{eqnarray}
where $\theta_n^{\pm}(k_x, k_y, \Lambda)$ is a $U(1)$ phase.

\begin{figure}
\includegraphics[width=7cm]{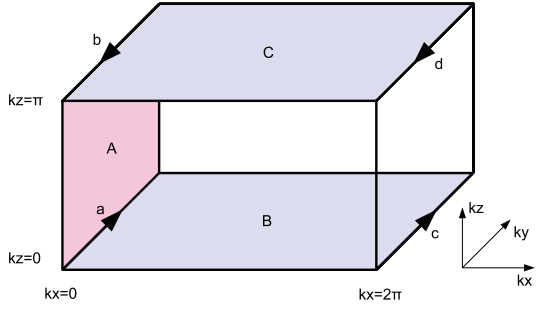} 
\caption{Upper half Brillouin zone.}
\label{sfig:hBZ}
\end{figure}

Now we consider the upper half region of the Brillouin zone in
Fig.\ref{sfig:hBZ}. 
From the twisted boundary condition,
the Berry phases $\gamma^{\pm}(\ell)$ along $\ell=a,b,c,d$ in
Fig.\ref{sfig:hBZ},  
\begin{eqnarray}
\gamma^{\pm}(\ell)=\oint_{\ell}dk_y A_y^{\pm}(k_x,k_y, \Lambda)
\end{eqnarray}
satisfy
\begin{eqnarray}
\gamma^+(a)=\gamma^-(c)\,\, (\mbox{mod. 2$\pi$}),  
\nonumber\\
\gamma^+(b)=\gamma^-(d)\,\, (\mbox{mod. 2$\pi$}).  
\label{seq:+-relation}
\end{eqnarray}
The Stokes's theorem also leads to
\begin{eqnarray}
&&\gamma(a+b)
=\int_A F_{yz}dk_y dk_z \,\, (\mbox{mod. 2$\pi$}),  
\nonumber\\
&&\gamma^{\pm}(c-a)
=\int_B F^{\pm}_{xy}dk_x dk_y \,\, (\mbox{mod. 2$\pi$}),  
\nonumber\\
&&\gamma^{\pm}(b-d)
=\int_C F^{\pm}_{xy}dk_x dk_y \,\, (\mbox{mod. 2$\pi$}),  
\label{seq:Berryrelation}
\end{eqnarray}
with $\gamma(\ell)=\gamma^{+}(\ell)+\gamma^-(\ell)$,
$F_{yz}=\partial_{k_y}A_z-\partial_{k_y}A_z$, and 
$F_{xy}^{\pm}=\partial_{k_x}A^{\pm}_y-\partial_{k_y}A^{\pm}_x$.
The modular equality in the above equations comes from the ambiguity
of the Berry phases. 


Using these relations, we find that the following $\nu$ defines 
the $\bm{Z}_2$ invariant
$(-1)^\nu$:
\begin{eqnarray}
\nu&=&\frac{1}{2\pi}\left[
\int_A F_{yz}dk_ydk_z
+\int_{B-C} F^{-}_{xy}dk_xdk_y
\right]
\nonumber\\
&-&\frac{1}{\pi}\gamma^+(a+b) \,\,(\mbox{mod. 2}). 
\label{seq:3dZ2}
\end{eqnarray}
Here note that the modulo-2 ambiguity from the Berry phase
$\gamma^{+}(a+b)$ does not affect on the $\bm{Z}_2$ invariant $(-1)^\nu$. 
In order for $(-1)^{\nu}$ to define the $\bm{Z}_2$ invariant, $\nu$
must be an integer. 
From Eqs.({\ref{seq:Berryrelation}) and (\ref{seq:+-relation}), we find
that
\begin{eqnarray}
&&\frac{1}{2\pi}\int_{B-C} F^{-}_{xy}dk_xdk_y
\nonumber\\
&&=\frac{1}{2\pi}\gamma^{-}(c-a-b+d)
\nonumber\\
&&=\frac{1}{2\pi}\left[
\gamma^{-}(c+d)-\gamma^{-}(a+b)
\right]
\nonumber\\
&&=\frac{1}{2\pi}\left[
\gamma^{+}(a+b)-\gamma^{-}(a+b)
\right]\,\,(\mbox{mod. 1}). 
\end{eqnarray}
Therefore, $\nu$ is recast into
\begin{eqnarray}
\nu&=&\frac{1}{2\pi}
\left[
\int_A F_{yz}dk_ydk_z-\gamma(a+b)
\right]\,\,(\mbox{mod. 1}), 
\end{eqnarray}
which takes an integer.

From the formula (\ref{seq:3dZ2}), we can calculate the ${\bm Z}_2$
invariant $(-1)^{\nu}$ for $H_{\rm 3D}({\bm k})$ in
Eq.(\ref{eq:3DNSHamiltonian}).
Since the ${\bm Z}_2$ invariant takes the same value unless the gap of the
system closes, we can choose the special case of $H_{\rm
1D}(k_x)=\sigma_z$. In this case, the first and the second terms of the
right hand side of Eq.(\ref{seq:3dZ2}) vanish, and thus we only need to
evaluate $\gamma^{+}(a+b)$.
On the glide invariant plane at $k_z=\Lambda$, $H_{\rm 3D}(k_x,
k_y,\Lambda)$ is decomposed into $H_{\rm 3D}^{\pm}$ in
the sector with the eigenvalue $\pm
e^{-ik_x/2}$ of $G(k_x)$,  
\begin{eqnarray}
&&H_{\rm 3D}^{\pm}(k_x,k_y,\Lambda)
\nonumber\\
&&=\pm
\begin{pmatrix}
\sin k_y & i(m+\cos \Lambda+\cos k_y) \\
-i(m+\cos \Lambda+\cos k_y) & -\sin k_y 
\end{pmatrix}.
\nonumber\\
\end{eqnarray}
From this, we find that
\begin{eqnarray}
&&\gamma^{+}(a)=\left\{
\begin{array}{ll}
0, 
& \mbox{for $m<-2$}\\
\pi, 
&\mbox{for $-2<m<0$}\\
0,
& \mbox{for $m>0$}
\end{array}
\right. ,
\nonumber\\
&&\gamma^{+}(b)=\left\{
\begin{array}{ll}
0, 
& \mbox{for $m<0$}\\
\pi, 
&\mbox{for $0<m<2$}\\
0,
& \mbox{for $m>2$}
\end{array}
\right. ,
\end{eqnarray}
which implies 
\begin{eqnarray}
\nu=\left\{
\begin{array}{ll}
0, 
& \mbox{for $m<-2$}\\
1, 
&\mbox{for $-2<m<0$}\\
1, 
&\mbox{for $0<m<2$}\\
0,
& \mbox{for $m>2$}
\end{array}
\right., 
\end{eqnarray}
modulo 2.
Therefore, $H_{\rm 3D}({\bm k})$ in Eq.(\ref{eq:3DNSHamiltonian}) is
topologically non-trivial (trivial) if
$-2<m<0$ or $0<m<2$ ($m>1$ or $m<-1$).

\section{K-theory analysis}

We summarize relevant results in the $K$-theory. 
Consider a class of nonsymmorphic symmetries, $\{U| \tau_x\}$, which
consist of a point group operation $U$
accompanying
a half translation $\tau_x$ of the lattice spacing in the $x$-direction.
We assume that the point group operation $U$ is a
$\mathbb{Z}_2$-transformation (namely order-two).
The nonsymmorphic symmetry $\{U|\tau_x\}$ acts on the Bloch Hamiltonian
$H({\bm k})$ as a $k_x$-dependent unitary transformation $U(k_x)$ with
$U^2(k_x)=e^{-ik_x/2}$.

Let us denote the K-group for $d$-dimensional insulators with the
nonsymmorphic symmetry $U(k_x)$ as $K_{\mathbb{Z}_2}^{(s,t,\tau_x)}(T^d)$. 
Here the superscript $(s,t,\tau_x)$ identifies the symmetries of the
insulators: $s=0, 1$ (mod.$2$) indicates the absence ($s=0$) or
the presence ($s=1$) of the additional chiral symmetry. 
Then $t=0,1$ (mod.$2$) determines how the point group operation of
$U(k_x)$ acts;
for $s=0$, $t$ specifies the action of $U(k_x)$ as \cite{Shiozaki2014}, 
\begin{eqnarray}
U(k_x) H(k_x,\bm{k}) U(k_x)^{-1} = 
\left\{ \begin{array}{ll}
H(k_x,\tilde{\bm{k}}) & (t=0), \\
-H(k_x,\tilde{\bm{k}}) & (t=1), \\
\end{array} \right.
\end{eqnarray}
and for $s=1$,
\begin{eqnarray}
&&\{H(k_x,\bm{k}), \Gamma\} = 0, 
\quad U(k_x) H(k_x,\bm{k}) U(k_x)^{-1} = H(k_x,\tilde{\bm{k}}), 
\nonumber\\
&&\Gamma U(k_x) = 
\left\{ \begin{array}{ll}
U(k_x) \Gamma & (t=0), \\
-U(k_x) \Gamma & (t=1), 
\end{array} \right.,
\end{eqnarray}
where $\bm{k} \mapsto \tilde{\bm{k}}$ represents a $\mathbb{Z}_2$
transformation for $T^{d-1}$.  
Finally, 
$\tau_x$ represents the half translation in the $x$-direction, as
mentioned above. 
%
$(t,\tau_x)$ is an example of a twisting of the twisted equivariant $K$-theory for topological insulators and superconductors. \cite{Freed2013}

From the Gysin exact sequence \cite{Karoubi2008} in the $K$-theory 
(the twisted version follows from the Thom isomorphism theorem \cite{Freed2011}), 
for $S^1$ except in the $k_x$-direction, 
the following isomorphism can be shown, 
\begin{eqnarray}
&&K_{\mathbb{Z}_2}^{(s,t,\tau_x)}(T^d \times S^1)
\nonumber\\ 
&&\cong K_{\mathbb{Z}_2}^{(s,t,\tau_x)}(T^d)
\oplus K_{\mathbb{Z}_2}^{(s-1,t,\tau_x)}(T^d),  
\label{seq:DimensionlReduction}
\end{eqnarray}
if $U(k_x)$ for $K_{\mathbb{Z}_2}^{(s,t,\tau_x)}(T^d \times S^1)$ acts
on $S^1$ as a global symmetry, or
\begin{eqnarray}
&&K_{\mathbb{Z}_2}^{(s,t,\tau_x)}(T^d \times S^1)
\nonumber\\ 
&&\cong K_{\mathbb{Z}_2}^{(s,t,\tau_x)}(T^d)
\oplus K_{\mathbb{Z}_2}^{(s-1,t-1,\tau_x)}(T^d),  
\label{seq:DimensionlReduction2}
\end{eqnarray}
if $U(k_x)$ for $K_{\mathbb{Z}_2}^{(s,t,\tau_x)}(T^d \times S^1)$
acts on $S^1$ as a reflection symmetry.
By iterating Eqs. (\ref{seq:DimensionlReduction}) and
(\ref{seq:DimensionlReduction2}), 
any K-group in the present case reduces to that for 
a $1$-dimensional nonsymmorphic insulator defined in the
$k_x$-direction.  
The Hamiltonian mapping $d=1 \to d=2 \to d=3$ discussed previously
is based on the isomorphism (\ref{seq:DimensionlReduction}) and  (\ref{seq:DimensionlReduction2}).
%
%
%
In Eq. (\ref{seq:DimensionlReduction}) or (\ref{seq:DimensionlReduction2}), 
the first term $K_{\mathbb{Z}_2}^{(n,t,\tau_x)}(T^d)$ in the right hand
side represents a ``weak" topological index of the left hand side, which is obtained
by just neglecting the $S^1$-dependence in the left hand side, 
but the second term gives the ``strong" topological index.

\end{document}